\documentclass[twocolumn, pra,showpacs,nofootinbib]{revtex4}
\usepackage{dcolumn,bm,graphicx}
\usepackage{amsmath}
\usepackage{amsfonts}
\usepackage{amssymb}

\begin{document}
\preprint{UNR Oct 2003-\today }
\title{Quantum computing with magnetically interacting atoms}
\author{Andrei Derevianko}
\author{Caleb C. Cannon }
\affiliation { Department of Physics, University of Nevada, Reno,
Nevada 89557}

\date{\today}

\begin{abstract}
We propose a scalable quantum-computing architecture based on
cold atoms confined to sites of a tight optical lattice.
The lattice is placed in a non-uniform magnetic field and the resulting
Zeeman sublevels define qubit states. Microwave
pulses tuned to space-dependent resonant frequencies
are used for individual addressing. The atoms interact
via magnetic-dipole interactions allowing implementation of a
universal controlled-NOT gate. The
resulting gate operation times for alkalis are on the
order of milliseconds,
much faster then the anticipated decoherence times. Single qubit operations
take about 10 microseconds. Analysis of motional decoherence due to
NOT operations is given.
We also comment on the improved feasibility of the proposed architecture
with complex open-shell atoms, such as Cr, Eu and metastable
alkaline-earth atoms with larger magnetic moments.
\end{abstract}


\maketitle

\section{Introduction}

Over the last decade, a variety of architectures for quantum computing (QC)
has been proposed~\cite{QCexp00}. In particular there is a number
of proposals based on neutral atoms and molecules trapped in optical lattices. These
proposals focus on various
realizations of the multi-qubit logic such as Rydberg-atom gates~\cite{JakCirZol00},
controlled collisions~\cite{JakBriCir99,YouCha00,BreDeuWil02}, electrostatic
interaction of heteronuclear molecules~\cite{DeM02}, etc (see a popular review~\cite{CirZol04}.)
While there is  a variety of approaches available,
the technological difficulties so far impede practical implementation
of these schemes.
Here we propose a scalable quantum-computing architecture
which further builds upon the well-established techniques of atom trapping.
Compared to the popular neutral-atom QC scheme with Rydberg gates~\cite{JakCirZol00},
the distinct features of the present
proposal are: (i) individual addressing of atoms confined to sites of 1D optical lattice
with {\em unfocused} beams of microwave radiation,
(ii) coherent ``always-on'' magnetic-dipolar interactions between the atoms,
and (iii) substantial decoupling of the motional and inner degrees of freedom.
The Hamiltonian of our system is identical to that of the QC based
on nuclear magnetic resonance (NMR)~\cite{LafKniCor02etal},
and already designed algorithms can be adopted for carrying computations
with our quantum processor. An implementation of the celebrated
Shor's algorithm with a linear array of qubits has been
discussed recently\cite{FowDevHol04}.

One of the challenges in choosing the physical system suitable for QC
is the strength of the interparticle interaction. Before proceeding
with the main discussion, let us elaborate on the suitability
of magnetic-dipolar atom-atom interactions for QC.
Compared to the dominant electrostatic interaction between a pair of atoms,
magnetic-dipole interaction between a pair
of atoms is weak (it is suppressed by a relativistic factor of $\alpha^2 \approx (1/137)^2$.)
Yet the dipolar interaction exhibits a pronounced anisotropic character:
the strength and the sign of the interaction depend on
a mutual orientation of magnetic moments of the two atoms.
Namely the {\em anisotropy} of the interaction plays a decisive
role in realizing a universal element of quantum logic: the
two-qubit controlled-NOT (CNOT) gate~\cite{BarDeuEke95}.
As to the {\em strength} of the interaction,
it determines how fast the two-qubit
gate operates. As the interaction strength decreases, the operation
time, $\tau_\mathrm{CNOT}$, increases.
Still, quantitatively, the operation of the gate
must be much faster than decoherence. For atoms
in far-off-resonance optical lattices, the decoherence
times for internal (hyperfine) states, chosen as
qubit states, are anticipated to be in the order of
minutes~\cite{qist}. On the other hand, we show that
for magnetically-interacting
atoms placed in tight optical lattices
$\tau_\mathrm{CNOT}$ is a few milli-second long.
Thus, although the interaction is weak, it is still strong
enough to allow for more than $10^4$ operations on a pair of qubits.
Considering that trapping millions of atoms is common now, the scalable
quantum computer (QC) proposed
here may present a competitive alternative to other architectures.

According to \citet{Div00}, the physical
implementation of QC
should satisfy the following five criteria:
(i) A scalable physical system with a well-characterized qubit;
(ii) The ability to initialize the state of the qubit to a simple
fiducial state;
(iii) A ``universal'' set of quantum gates;
(iv) Long relevant decoherence times, much longer than the gate
operation time;
(v) A qubit-specific measurement capability.
Below, while describing our proposed QC, we explicitly address these criteria.
We also compare it with QC
based on nuclear magnetic resonance (NMR)
techniques~\cite{LafKniCor02etal}, because
of the equivalence of multiparticle Hamiltonians of our QC and the NMR systems.
Unless noted otherwise, atomic units $|e|=m_e=\hbar\equiv 1$ are used
throughout; in these units the Bohr magneton $\mu_B=1/2$.

\section{Choice of qubit}
\begin{figure}[h]
\begin{center}
\includegraphics*[scale=0.75]{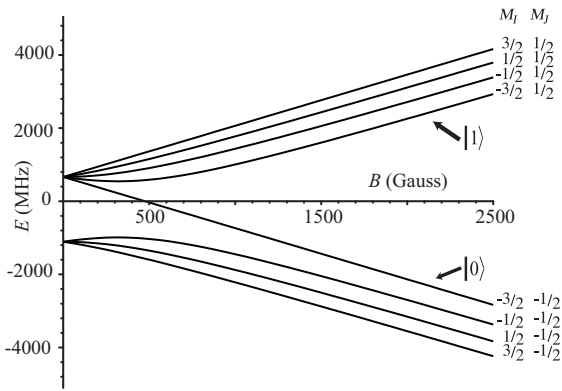}
\caption{ \label{Fig:ZeemanHFS}
 Zeeman effect for the ground state of $^{23}$Na atom.
 The states are labeled with the electronic and nuclear magnetic quantum numbers
 $M_J$ and $M_I$ in the strong-field limit. A possible choice of qubit
 states is shown.
 }
\end{center}
\end{figure}

We focus on alkali-metal atoms in their respective ground $^2\!S_{1/2}$ state,
but later (Section~\ref{Sec:ComplexAtoms}) also comment on the improved feasibility of our QC for
other open-shell ground state and metastable atoms.
The atoms are placed in a magnetic field and the resulting Zeeman components
define qubit states. As shown in Fig.~\ref{Fig:ZeemanHFS}
for $^{23}$Na,
the Zeeman energies behave non-linearly as a function of the magnetic field
strength $B$. The non-linearity is due to an interplay between
couplings of atomic electrons to the nuclear and the externally-applied
fields.
The discussion presented below can be extended to an arbitrary field strength,
but for illustration we consider the high-field limit,
$\mu_B B \gg \Delta E_\mathrm{HFS}/(2I+1)$. Here
$\Delta E_\mathrm{HFS}$ is the hyperfine splitting that
ranges from 228 MHz for $^{6}$Li to
9193 MHz in $^{133}$Cs and $I$ is the nuclear spin.
In this regime, the proper lowest-order
states are product states $|J,M_J\rangle |I,M_I\rangle$, where $M_J$
and $M_I$ are projections of the total electronic momentum $J=1/2$ and
the nuclear spin $I$ along the B-field.
We choose the qubit states as
$|1\rangle \equiv |J,M_J=1/2\rangle |I,M_I\rangle$ and
$|0\rangle \equiv |J,M_J=-1/2\rangle |I,M_I\rangle$. Disregarding
nuclear moment, these states have magnetic
moments of $\mu_{|1\rangle}=\mu_B$ and $\mu_{|0\rangle}=-\mu_B$,
and the associated Zeeman splitting is on the order of a few GHz (see Fig.\ref{Fig:ZeemanHFS}).
This choice allows driving single-qubit unitary operations with microwave (MW) pulses.
The MW radiation has to be resonant with the Zeeman frequency
$\left(\mu_{|1\rangle} -\mu_{|0\rangle} \right) B$.

As to the initialization of the qubits of the proposed QC, the conventional
techniques of optical pumping may be used. State-selective ionization may
be employed for the read-out of the results of calculations. Ion
optics for registering the final states is described in Ref.~\cite{DeM02}.
Both the initialization process and the projective readout favor the
proposed QC in comparison to the liquid state NMR QC, where the ensemble
averaging is required for read-out and relaxation of the sample
is important for initialization.

\section{Individual addressability}
\begin{figure}[h]
\begin{center}
\includegraphics*[scale=0.4]{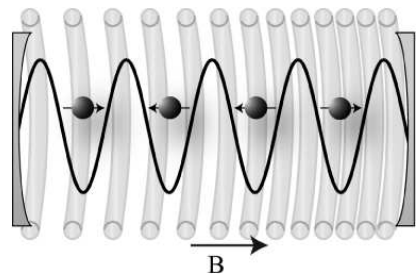}
\caption{ \label{Fig:Architecture}
Proposed quantum computing architecture. The atoms are confined
to sites of an optical lattice. Non-uniform magnetic field is
required for individual addressing of the atoms with pulses of the resonant
microwave radiation.
 }
\end{center}
\end{figure}

In NMR~\cite{LafKniCor02etal}, the nuclear spins (qubits) are distinguished by their different
chemical shifts affecting resonance spin-flip frequencies.
Here to individually address the atoms, we
confine ultracold atoms to sites of a one-dimensional optical
lattice and introduce a gradient of magnetic field,
so that the Zeeman frequency depends on the position of the atom
in the lattice (see Fig.\ref{Fig:Architecture}).
This addressing scheme was investigated in details in Ref.~\cite{MinWun01},
in the context of quantum processor using trapped ions.
It is also worth mentioning a similar idea for QC with heteronuclear
diatomic molecules~\cite{DeM02}, where a gradient of electric field
was applied along optical lattice affecting flip frequencies of molecular
dipoles.

\subsection{Optical lattice}

The 1D potential of the optical lattice
created by a standing wave CW laser beam
reads  $V_\mathrm{opt}(z)=
V_{0}\sin^{2}\left( \frac{2\pi}{\lambda}z\right)$.
The depth of the wells is
$V_0= 8 \pi\alpha \times \alpha_a(\omega) I_L$,
where $\alpha_a(\omega_L)$ is the dynamic
electric-dipole polarizability of the atom, $I_L$ is the laser intensity, and $\alpha \approx 1/137$
is the fine-structure constant.
Depending on the detuning of the
laser frequency $\omega_L$ from a position of atomic resonance,
the polarizability can accept a wide range of values.
To restrict the transverse atomic motion in a
1D optical lattice one must require that $\alpha_a(\omega_L)>0$.
The case  of negative $\alpha_a(\omega_L)$, although
requiring 3D optical lattice, offers an advantage
of reduced photon scattering rates. Since the atoms
would be located at the intensity minima, the rates
(in a tight confinement regime) would be suppressed by a factor
of $1/2 \sqrt{E_R/V_0} \ll 1$, where $E_R = \alpha^2 \omega_L^2 /(2 M)$
is the photon recoil energy for an atom of mass $M$.
Further, an atom is assumed to occupy the ground motional state of the lattice wells
and
one atom per site filling ratio is assumed. Techniques for loading optical
lattices are being perfected~\cite{latticeRef,BrePupRey03}.

In the 1D optical lattice,
the neighboring atoms are separated by a distance of
$R=\lambda/2$.
To maximize the magnetic-dipole interaction ($\propto 1/R^3$) between two
neighbors we require that the frequency of the laser $\omega_L$
is chosen as high as possible. A natural limit on $\omega_L$
is the ionization potential (IP) which ranges from 3.8 eV for Cs to 5.4 eV for Li.
Practically, $\omega_L$ must be somewhat below  the IP to avoid resonances in the high-density
of states near the continuum limit.
In the estimates below we use $\omega_L \sim 5$ eV. This corresponds
to
\[
\lambda_L \approx 250 \, \mathrm{nm} = 4700 \, \text{bohr} \,.
\]
While working
with such short ultra-violet wavelengths seems feasible, more conventional wavelenghts of
$400-600 \mathrm{nm}$ are adequate when working with complex open-shell
atoms with larger magnetic moments (see Section~\ref{Sec:ComplexAtoms}).

One additional requirement
is that the tunneling time between adjacent sites of an optical lattice
is sufficiently long, so that the atoms maintain their distinguishability
during the computation. The site-hoping frequency can be estimated as~\cite{BrePupRey03}
\[
J=E_{R}\frac{4}{\sqrt{\pi }}
\left( \frac{V_{0}}{E_{R}}\right) ^{3/4}\exp \left[ -2\sqrt{V_{0}/E_{R}}\right].
\]
For $^{23}\mathrm{Na}$, the requirement that the characteristic tunneling time $\pi/J\approx
10 \, \mathrm{sec}$
leads to a barrier hight of $V_0=0.4 \, \mathrm{mK}$.
At a fixed $\omega_L$, the tunneling may be suppressed
by increasing the barrier height $V_0$ and by using heavier atoms.
Additionally, due to the Pauli exclusion principle,
the tunneling may be suppressed by using fermionic atoms.

\subsection{B-field gradients}
\label{Sec:dBdz}
The magnetic field leads to the field-dependent Zeeman effect and
the {\em gradient} $dB/dz$ of the B-field allows one to resolve the resonant frequencies
of individual atoms, see Fig.~\ref{Fig:Architecture}. In the Section~\ref{Sec:CNOT},
we show that a typical two-qubit operation has a duration $\tau_\mathrm{CNOT}$ of
a few milliseconds. To find the gradient $dB/dz$ we require that
the single-qubit NOT operation, performed by the resonant $\pi$-pulse of microwave radiation,
takes a comparable time $\tau_\mathrm{NOT}$ of 1 ms. Such pulse may be resolved by two neighboring atoms,
provided that their resonant frequencies  differ by
$\Delta f_\mathrm{NOT} =1/(2 \pi \tau_\mathrm{NOT}) \approx 2 \times 10^{2} \, \mathrm{Hz}$.
The required field gradient is relatively weak
\[
dB/dz  \approx 20 \, \mathrm{G/cm}
\]
and is comparable to typical gradients in conventional magnetic traps.
Much steeper gradients of $3 \times 10^3 \, \mathrm{G/cm}$ over a
region of a few millimeters have been demonstrated by \citet{VulHanZim96}.
These authors employed ferromagnetic needles that collect and focus B-fields of
electromagnetic coils. With such gradients the performance of the single-
qubit gates can be substantially improved, $\tau_\mathrm{NOT} \sim  10 \, \mu \mathrm{s}$.

Another limitation on the gradients is that the magnetic force $\mu dB/dz$
must be much smaller than the optical force $-dV_\mathrm{opt}(z)/dz$.
This limitation affects not only the confinement but also a degree of coupling
of inner and motional degrees of freedom because of the difference in
values of magnetic moments for the two qubit states (see Section~\ref{Sec:InnerMotional}).
For the parameters chosen above the magnetic forces are several orders of magnitude
smaller than the optical ones.

\section{Multi-qubit operations}
\label{Sec:CNOT}
Having discussed one-qubit operations, we now
consider a realization of the universal
controlled-NOT gate~\cite{BarDeuEke95} based on magnetic-dipolar
interaction of two atoms. It is worth mentioning that we discuss
this gate here only for illustrative purposes, to estimate the characteristic duration
of the CNOT operation,
$\tau_\mathrm{CNOT}$. The many-body dynamics of the system with the interparticle interaction
which is ``always on'' is such that the two-qubit gates are executed
as a part of the natural dynamics of the system, and a special care
has to taken to govern this natural evolution~\cite{LeuChuYam00}. We will elaborate
on this point at the end of this Section.

The basic requirement for the controlled-NOT gate~\cite{BarDeuEke95}
is that the frequency of $|1\rangle \rightarrow |0\rangle$
transition of the target atom depends on the state of the control atom.
Since the quantization (B-field) axis is
directed along the internuclear separation between the two atoms,
the magnetic-dipole interaction can be represented  as
\begin{equation}
 V = -\frac{\alpha^2}{R^3} \sum_{\lambda=-1}^{1} (1+\delta_{\lambda,0})
 \mu_\lambda(1) \mu_{-\lambda}(2) \, ,
 \label{Eq:VmuXmu}
\end{equation}
where $\mu_\lambda(k)$ represents the spherical components
of the magnetic moment operator $\bm{\mu}$
for atom $k$.
%
%
For performing the CNOT gate operation we need to resolve the frequency
difference of $2\alpha^{2}R^{-3} \left(\mu_{|1\rangle}-\mu_{|0\rangle}\right)^{2}$.
For our choice of parameters we find that
$\Delta f_\mathrm{CNOT} \approx 40 \, \mathrm{Hz}$.
Although this number may seem small, it is comparable to typical coupling
strengths of 20-200 Hz in NMR. Furthermore, the minimum duration of the MW pulse is
$\tau_\mathrm{CNOT}=1/(2\pi\Delta f_\mathrm{CNOT}) \approx 3 \, \mathrm{ms}$,
allowing for more than $10^4$ CNOT operations during anticipated
decoherence time (in the order of minutes~\cite{qist}).

As in the NMR, the interaction between the atoms is always on and special care
has to be taken to stop the undesired time evolution of the system and
carry out controlled calculations. Fortunately, it is straightforward to
show that the many-particle Hamiltonian of our system is equivalent to that of a
system of interacting spins in NMR and thus already developed techniques
can be adopted. Specifically, the Hamiltonian is
\begin{equation}
H_\mathrm{NMR}=\sum_{i}\omega_{i}\left(  \sigma_{z}\right)  _{i}+\frac{1}%
{2}\sum_{ij}g_{ij}\left(  \sigma_{z}\right)  _{i}\left(  \sigma
_{z}\right)  _{j}. \label{Eq:HNMR}
\end{equation}
Here $\left(  \sigma_{z}\right)_{i}$ are the Pauli matrices for an atom $i$
located at position $z_i$. Introducing the average magnetic
moment
$\bar{\mu} = (\mu_{|1\rangle} +\mu_{|0\rangle})/2$ and
the difference $\delta\mu=(\mu_{|1\rangle} -\mu_{|0\rangle})/2$,
the one-particle frequencies may be expressed as
\[
\omega_{i}= -\delta\mu_{i}B\left(
z_{i}\right)   +\sum_{j}\frac{2\alpha^{2}}{|z_{i}-z_{j}|^{3}}\bar{\mu
}_{j}\delta\mu_{i}
\]
and the two-body couplings as
\[
g_{ij}=-2\alpha^{2}/(|z_{i}-z_{j}|^{3}) \, \delta\mu_{i}\delta\mu_{j} \, .
\]

Since the Hamiltonian $H_\mathrm{NMR}$, Eq.(\ref{Eq:HNMR}), is
identical to that of the NMR based QC~\cite{LeuChuYam00},
the already developed NMR algorithms can be adopted.
For example, the time evolution
for a given atom may be reversed by inverting populations.
The same idea is applicable for  the interparticle couplings.
The CNOT gate may be implemented by applying a sequence of one-qubit transformations
(short pulses) and allowing a given coupling to develop in
time~\cite{LeuChuYam00}. An implementation of the celebrated
Shor's algorithm on the system of linear array of qubits has been
discussed recently by \citet{FowDevHol04}.

An important point is that the two-particle couplings result
in a {\em coherent} development
of the system. As we demonstrate below, the short NOT pulses
introduce negligible decoherences due to excitation of motional
quanta during the gate operation.
\section{Decoherences}

Finally, we address possible sources of decoherences. We divide
these sources into two classes: (i) induced due to the required operation
of the quantum processor (due to performing gates) and (ii) architectural,
i.e., due to traditional sources of decoherences, such as
photon scattering (e.g. Raman), instabilities of laser output, instabilities in the
magnetic field, collisions with background gas, etc.
Below we analyze the decoherences caused by the gate operations and
we also estimate the upper limit on tolerable fluctuations of the magnetic field.
The remaining decoherences depend on the  details of experimental design, e.g.,
laser wavelengths available, and we can not fully address them at this point.

\subsection{Motional heating}
\label{Sec:InnerMotional}
The atomic center-of-mass (C.M.)
evolves in a combined potential of the optical lattice
and magnetic field. This potential depends on the
internal state of the atom, due to a difference in
dynamic polarizabilities and magnetic moments of the states.
When population is transferred from one qubit (internal) state to the other,
the motion of the C.M. is perturbed and the atom may leave the ground state of
C.M. potential (motional heating). Here we calculate the relevant probability and show that it is
negligible. We find that the motional heating is suppressed due to adiabaticity of
the population-transfer process. In other words, the inner and motional degrees of
freedom decouple because our quantum processor is relatively slow.
It is worth emphasizing  that in the popular neutral-atom QC proposal~\cite{JakCirZol00},
the operation of the gates is much faster and the motional heating is of concern~\cite{SafWilCla03}.

As discussed in the previous section, the QC proposed here requires only
single-qubit rotation pulses with a characteristic duration $\tau_\mathrm{NOT}$.
The two-qubit gate operations are performed by the NOT pulses and due to the
natural coherent dynamics of the system, so it sufficient to consider the
coupling of the inner and motional degrees of freedom due to the NOT pulse only.

For a resonant radiation, during the MW pulse, the effective Hamiltonian may be represented as
\begin{eqnarray}
H(\mathbf{R},\mathbf{r}) &=&
\frac{\Omega}{2}\left(  |0\rangle\langle 1|+|1\rangle\langle 0|\right)+ \\
& &\frac{P^{2}}{2M}+V_{|1\rangle}\left( \mathbf{R}\right)  |1\rangle\langle 1|+
V_{|0\rangle}\left(\mathbf{R}\right)  |0\rangle\langle 0|. \nonumber
\end{eqnarray}
Here $\mathbf{R}$ and $\mathbf{P}$ are the C.M. coordinate and
momentum, and $\mathbf{r}$ encapsulates internal degrees of freedom
($|0\rangle$ and $|1\rangle$).
The internal states are
coupled via $\Omega=\pi/\tau_\mathrm{NOT}$, the Rabi frequency of
the MW transition. The C.M. evolves either in $V_{|0\rangle}$ or
$V_{|1\rangle}$ potentials. We treat
the difference between the C.M. potentials
$W=V_{|1\rangle}-V_{|0\rangle}$ as a perturbation and characterize
the C.M. motion with eigenstates $|\Phi_n\rangle$ and
energies $\mathcal{E}_n$ computed in the $V_{|0\rangle}$ potential.
\[
\left( \frac{P^{2}}{2M}+
V_{|0\rangle}\left(\mathbf{R}\right) \right)|\Phi_n\rangle =\mathcal{E}_n |\Phi_n\rangle \, .
\]
We assume that at $t=0$ the atom is in the motional and
internal ground state
$\Psi(\mathbf{R},\mathbf{r}, t=0)=|0\rangle \, |\Phi_0\rangle$.
As a result of the transfer of
population (NOT operation, $|0\rangle \rightarrow |1\rangle$ )
there is a finite probability $P_p$
for an atom to end up in one of the excited motional
eigenstates $|\Phi_p\rangle$.

\begin{figure}[h]
\begin{center}
\includegraphics*[scale=0.5]{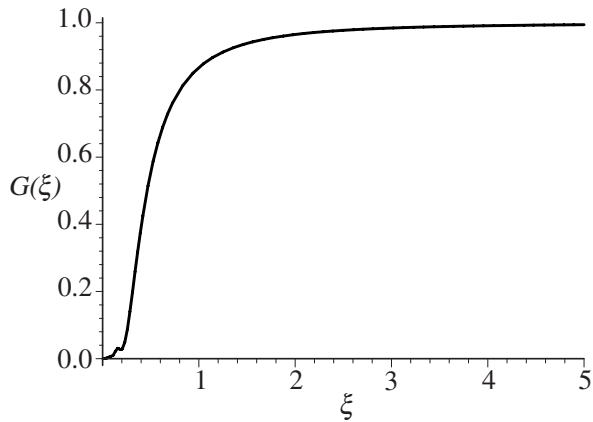}
\caption{ \label{Fig:Gxi}
Adiabaticity function $G\left(  \xi \right)$, Eq.(\protect\ref{Eq:Gxi}). }
\end{center}
\end{figure}

We evaluate the probability $P_p$ perturbatively  by expanding
\[
\Psi(\mathbf{R},\mathbf{r}, t) \approx a(t)|0\rangle \, |\Phi_0\rangle +
b(t)|1\rangle \, |\Phi_0\rangle + c(t)|1\rangle \, |\Phi_p\rangle .
\]
Substituting this expansion into the Schrodinger equation and
solving it perturbatively (perturbation is $W$, the two-level  Rabi evolution is treated exactly )
we obtain
\begin{eqnarray*}
a(t)& \approx &\sin(\Omega/2 t) \, ,\\
b(t)& \approx &\cos(\Omega/2 t) \, ,\\
i\frac{d}{dt}c \left( t\right)
& \approx & \sin \left(\Omega/2 t \right) \exp \left( i \omega _{p0}t\right)
\langle\Phi_p| W |\Phi_0 \rangle \, ,
\end{eqnarray*}
where $\omega _{p0} = \mathcal{E}_p - \mathcal{E}_0 $.
Calculation of the probability based on $c(t)$ requires certain care, since after the initial
application of the pulse the perturbation $W$ remains on indefinitely. Using
an approach discussed in Ref.~\cite{LanLif97}, we arrive at the probability of excitation
\begin{equation}
P_p(\tau_\mathrm{NOT}) = \left\vert \frac{\langle\Phi_p| W |\Phi_0 \rangle }
{\omega _{p0}}\right\vert^{2}
G\left( \frac{\Omega/2}{\omega _{p0}} \right),
\label{Eq:Pmotional}
\end{equation}
where we introduced the adiabaticity function
\begin{equation}
G\left(  \xi \right)  =\frac{\xi^{2}%
}{\left(  \xi^{2}-1\right)  ^{2}}\,\left(  1+\xi^{2}-2\,\xi\,\sin(\frac{\pi
}{2\,\xi})\right) \, .
\label{Eq:Gxi}
\end{equation}
This function is plotted in Fig.~\ref{Fig:Gxi}.
The function is bounded $G\left(  \xi \right) \le 1$;
there is no
singularity at the resonant frequency $G(\xi=1)\approx 0.87$, since the
C.M. experiences only a quarter of the period of oscillating function
during the $\pi$ pulse.
Let us investigate various limits
of $G\left(  \xi \right)$.
For an instantaneous turn-on of the perturbation, $G(\xi \gg 1) \approx 1$, so that
\[
P_p(\tau_\mathrm{NOT}  ) = \left\vert \frac{\langle\Phi_p| W |\Phi_0 \rangle }
{\omega _{p0}}\right\vert^{2}, \, \tau_\mathrm{NOT}\omega _{p0} \ll 1 \,.
\]
For a slow (adiabatic) application of the $\pi$-pulse
$G(\xi \ll 1) \approx 1/\xi^2$. In this adiabatic limit
\[
P_p(\tau_\mathrm{NOT} ) = \left\vert \frac{\langle\Phi_p| W |\Phi_0 \rangle }
{\Omega/2}\right\vert^{2} \, , \tau_\mathrm{NOT}\omega _{p0}  \gg 1 \,.
\]

Now we can carry out the numerical estimates. To compute
the matrix elements of the perturbation, we approximate the bottom of the optical
potential as  that of  harmonic oscillator of the frequency
$\omega_\mathrm{ho}=\frac{2\pi}{\lambda_L}\sqrt{2V_{0}/M}$ and use
the harmonic oscillator wavefunctions for $|\Phi_n\rangle$. For
$^{23}$Na and our lattice parameters ($\lambda_L\approx 250 \, \mathrm{nm},
V_0 \approx 0.4 \, \mathrm{mK}$ ), $\omega_\mathrm{ho} \approx
2\pi~\times~2 \times 10^{6} \, \mathrm{Hz}$, while the Rabi frequency
(for the extreme case of $dB/dz = 2 \times 10^3 \mathrm{G/cm}$)
$\Omega/2 = 2 \pi \times 0.25 \times 10^5 \, \mathrm{Hz}$, i.e. even in the case of steep gradients
we deal  with an adiabatically slow pulse and the motional heating is suppressed
by a factor of $\approx [\omega_{p0}/(\Omega/2)]^2$.

For a perturbation due to
an interaction of magnetic moments with the gradient of magnetic field
$W = \left(\mu_{|0\rangle} -\mu_{|1\rangle}\right) (dB/dz) z$.
For our parameters, the probability of exciting motional quantum
is just $P_1 \sim 6 \times 10^{-7}$.

Another
source of coupling of internal and motional degrees of freedom is due to
the difference in the dynamic
polarizability $\alpha_a(\omega)$ for the two qubit states. This difference
leads to a modification
of $\omega_\mathrm{ho} \propto \sqrt{\alpha_a(\omega)}$. The relevant
probability may be estimated as $P_2 \approx
G\left(  \Omega/4\omega_\mathrm{ho}\right)
\frac{1}{32}\left(  2 \mu_B B /\Delta\omega\right)^{2}
\approx 10^{-10} \left(   2 \mu_B B /\Delta\omega\right)^{2}$,
where $\Delta \omega$  is a detuning of the laser frequency from the energy of
the intermediate state contributing the most to the dynamic polarizability
$\alpha_a(\omega_L)$. (Here we used the field gradient $dB/dz = 20 \mathrm{G/cm}$, so that
$\tau_\mathrm{NOT} \approx \tau_\mathrm{CNOT}$, see section \ref{Sec:dBdz}.)
In practice,  $ 2 \mu_B B /\Delta\omega \ll 1 $,
so that $P_2 \ll 10^{-10}$. This probability can be reduced even further by adjusting
the laser wavelength so that the dynamic polarizabilities of the two qubit states
are the same~\cite{SafWilCla03}.

Finally, we would like to emphasize that the motional heating is not
an issue for our processor, because our QC is relatively {\em slow}. The
C.M. cycles through many oscillations while the perturbing MW field is applied.
This adiabatic averaging leads to the suppression of the motional heating.

\subsection{Decoherences due to fluctuations of the magnetic field}
The atoms in our quantum processor are required to have magnetic moments.
This magnetic moments can couple to the ambient magnetic field. The fluctuations
of these ambient fields can cause the qubit states to lose coherence. In addition,
the Johnson/shot noise in the coils providing the gradient of the B-field can lead to the
decoherences as well. Here we estimate the upper limit of tolerance to these fluctuations.

Before we carry out the estimate, we notice that the sensitivity to fluctuations
of the magnetic fields can be avoided altogether using so-called
Decoherence Free Subspaces (DFS)~\cite{DuaGuo98,LidChuWha98,ZanRas97}.
The idea of the DFS  is to redefine the qubit states using linear combinations of
product of states of original qubits;
each resulting state accumulates the same phase due to environmental interactions. Since the
total phase of wavefunction is irrelevant, the DFS is highly stable with the respect to
the external perturbations.
This powerful DFS approach has been already
experimentally verified in a number of cases, including  trapped ions~\cite{KieMeyRow01}.
Nevertheless, an introduction of the DFS requires reconsideration of the NMR algorithms,
which is beyond the scope of the present paper. Below we estimate the decoherence of our
original qubit defined as a couple of magnetic substates of an atom.

To make an order-of-magnitude estimate, we assume that the fluctuating field $B'(t)$ is along
the quantization axis and it has a white noise spectrum. With the fluctuating
field present, the accumulated phase difference between the qubit states is
$\delta\phi \approx 2\int_{0}^{t}\mu_{|0\rangle}B^{\prime}\left(  t_{1}\right)  dt_{1}$.
(Here we assumed that $\mu_{|1\rangle} \approx -\mu_{|0\rangle}$.)
Averaging over different realizations
\begin{eqnarray*}
\left\langle \exp\left(  i\delta\phi\right)  \right\rangle & \approx &
1+i\left\langle \delta\phi\right\rangle -\frac{1}{2}\left\langle \delta
\phi^{2}\right\rangle \\
&= &1+2i\int_{0}^{t}\mu_{|0\rangle}\left\langle B^{\prime}\left(
t_{1}\right)  \right\rangle dt_{1}\\ &&-2\mu_{|0\rangle}^{2}\int_{0}^{t}\int_{0}%
^{t}\left\langle B^{\prime}\left(  t_{1}\right)  B^{\prime}\left(
t_{2}\right)  \right\rangle dt_{1}dt_{2} \, .
\end{eqnarray*}
This expression may be simplified using ensemble average $\left\langle B^{\prime}\left(  t_{1}\right)
\right\rangle =0$, and the autocorrelation function for a stochastic process with
no memory
\[
\left\langle B^{\prime}\left(  t_{1}\right)  B^{\prime}\left(  t_{2}\right)
\right\rangle =\left(  B^{2}\right)  _{\omega}\delta\left(  t_{1}%
-t_{2}\right)\, ,
\]
where $\left(  B^{2}\right)_{\omega} =\mathrm{const}$. If a measurement
is carried out after $\tau_\mathrm{CNOT}$, the probabilities would differ from the exact result by
\[
 \varepsilon_B = (\left\langle \exp\left(  i\delta\phi\right)  \right\rangle-1)^2 \approx
4 \mu _{|0\rangle}^{2}~\left( B^{2}\right) _{\omega } \tau_\mathrm{CNOT} \, .
\]
According to \citet{Kni04},
this error can be as high as 1\%. This leads to a
tolerable level of the B-field noise
\[
\sqrt{\left(  B^{2}\right)_{\omega}} \lesssim 10^{-10}\frac{\mathrm{T}}{\sqrt{\mathrm{Hz}}} \, .
\]
This limit is relatively easy to attain experimentally~\cite{RomPrivate}.

To summarize the main results of this section, we have demonstrated
that for our quantum processor the motional heating caused by the gate
operations is negligible.
The derived formula ~(\ref{Eq:Pmotional}) is also applicable for an
analysis of coupling of inner and motional
degrees of freedom of QC based on heteronuclear molecules~\cite{DeM02}.
We also have evaluated the tolerable level of fluctuations in magnetic field.
It is worth emphasizing that the sensitivity to environmental field noise
can be greatly eliminated using decoherence free subspaces, as discussed in
the beginning of this subsection.

There are other potential sources of decoherences, e.g.
photon scattering, instabilities of laser output, and collisions with background gas.
However, we do not address them
at this point since they depend on specifics of experimental design.
Still it is anticipated~\cite{qist} that the decoherence
times for internal states may be in the order of
minutes.

\section{Complex open-shell atoms}
\label{Sec:ComplexAtoms}
In this paper we focused on alkali-metal atoms ($\mu = \mu_B$) and we found that
wavelengths of $\lambda \sim  250$ nm are needed for
$\tau_\mathrm{CNOT} \approx 3 \, \mathrm{ms}$. While working
with such short ultra-violet wavelengths still seems feasible, the restriction
may be relaxed by
using complex open-shell atoms with larger values of magnetic moments.
This allows increasing the interaction strength ($\propto \mu^2$) and
thus working with more conventional laser wavelengths.

For example, recently
cooled and trapped Cr and Eu atoms\cite{CrEuRefs} have magnetic moments of stretched states
$\mu_\mathrm{Cr} = 6 \mu_B$ and $\mu_\mathrm{Eu} = 7 \mu_B$. Thus if a laser
of a 400 nm wavelength is employed for constructing the optical lattice,
the CNOT-gate operations would require
$\tau_\mathrm{CNOT} \sim 0.1 \,  \mathrm{ms}$. The use of 30-mW violet 400 nm
LD laser has been recently demonstrated in cooling and trapping
experiments~\cite{ParYoo03}.  The requirement
$\tau_\mathrm{CNOT} \sim 1 \, \mathrm{ms}$ would bring the wavelength
to an even more practical 800 nm.

 Another alternative is to use long-lived~\cite{Der01}
metastable divalent atoms, Mg, Ca, Sr and Yb. Several groups
have demonstrated cooling and trapping these
atoms~\cite{metaAErefs}, with
Yb BEC recently attained\cite{TakMakKom03}. Here the state of
interest is $^3\!P_2$, with $\mu_{^3\!P_2} = 3 \mu_B$ and for a 400 nm laser
$\tau_\mathrm{CNOT} \sim 1\,  \mathrm{ms}$.

\section{Conclusion}

In summary, we proposed a scalable quantum computing architecture
based on magnetically-interacting cold atoms confined to
sites of a tight optical lattice. The atoms
are placed in a non-uniform magnetic field and the individual
addressing is attained by pulses of microwave radiation for a minimum
duration of 10 $\mu$s.  The universal two-qubit CNOT gates require times in the order
of milliseconds. The multi-particle Hamiltonian of the system is equivalent
to that of the QC based on NMR techniques so the already developed
quantum algorithms may be adopted.

Compared to the popular neutral-atom QC scheme~\cite{JakCirZol00},
the distinct advantages of the present
proposal are: (i) individual addressability of atoms
with {\em unfocused} beams of microwave radiation,
and (ii) coherent evolution due to the ``always-on'' magnetic-dipolar interactions between the atoms,
(iii) substantial decoupling of the motional and inner degrees of freedom.
While the main disadvantage of our QC is the slow rate of operation,
it is anticipated that one could carry out
$10^4$ CNOT operations on a single pair of atoms before
the coherence is lost. When addressing tens of thousands of atoms
is done in parallel, this would amass $10^8$ two-qubit operations
and $10^{10}$ single-qubit operations.

\acknowledgements

We would like to thank  J. Thompson, T. Killian, E. Cornell, C. Williams, D. DeMille,
and M. Romalis for discussions,  and X. Tang  for
a review of NMR techniques.
This work was supported in part by the National Science Foundation.


\end{document}